\documentclass[aps,pre,superscriptaddress,amsmath,amssymb,amsfonts,twocolumn,showpacs]{revtex4-1}
\usepackage[utf8]{inputenc}
\usepackage{amsmath}
\usepackage{amssymb}
\usepackage{amsfonts}
\usepackage{amstext}
\usepackage{amsthm}
\usepackage{mathtools}
\usepackage{physics}
\usepackage{graphicx}
\usepackage{dcolumn}
\usepackage{bm}
\usepackage{braket}
\usepackage{color}
\usepackage[colorlinks=true,citecolor=blue]{hyperref}
\usepackage{natbib}
\usepackage{fourier,times}
\usepackage[english]{babel}
\usepackage{xcolor}
\usepackage{setspace}

\usepackage[normalem]{ulem} %% REMOVE ME!

\definecolor{deeppurple}{rgb}{0.7, 0, 0.8}

\allowdisplaybreaks
\advance\parskip 0.4pt

\begin{document}
\setstretch{1.08}

\title{Experimental realization of the Peregrine soliton in repulsive two-component\\ Bose-Einstein condensates}

\author{A. Romero-Ros}
\affiliation{Center for Optical Quantum Technologies, Department of Physics,
    University of Hamburg, Luruper Chaussee 149, 22761 Hamburg,	Germany}

\author{G. C. Katsimiga}
\affiliation{Department of Mathematics and Statistics, University of Massachusetts Amherst, Amherst, MA 01003-4515, USA}

\author{S. I. Mistakidis}
\affiliation{ITAMP, Center for Astrophysics $|$ Harvard $\&$ Smithsonian, Cambridge, MA 02138 USA}
\affiliation{Department of Physics, Harvard University, Cambridge, Massachusetts 02138, USA}

\author{S. Mossman}
\affiliation{Department of Physics and Biophysics, University of San Diego, San Diego, CA 92110}
\affiliation{Department of Physics and Astronomy, Washington State University, Pullman, Washington 99164-2814}

\author{G. Biondini}
\affiliation{Department of Mathematics, State University of New York, Buffalo, New York 14260, USA}
\affiliation{Department of Physics, State University of New York, Buffalo, New York 14260, USA}

\author{P. Schmelcher}
\affiliation{Center for Optical Quantum Technologies, Department of Physics,
    University of Hamburg, Luruper Chaussee 149, 22761 Hamburg,	Germany}
\affiliation{The Hamburg Centre for Ultrafast Imaging,
    University of Hamburg, Luruper Chaussee 149, 22761 Hamburg,	Germany}

\author{P. Engels}
\affiliation{Department of Physics and Astronomy, Washington State University, Pullman, Washington 99164-2814}

\author{P. G. Kevrekidis}
\affiliation{Department of Mathematics and Statistics, University of Massachusetts Amherst, Amherst, MA 01003-4515, USA}

\date{\today}

\begin{abstract}
    We experimentally realize the Peregrine soliton in a highly particle-imbalanced two-component repulsive Bose-Einstein condensate in the immiscible regime. 
    The effective focusing dynamics and resulting modulational instability of the minority component provide the opportunity to dynamically  create a Peregrine soliton with the aid of  an attractive potential well that seeds the initial dynamics. 
    The Peregrine soliton formation is highly reproducible, and our experiments allow us to separately monitor the minority and majority components, and to compare with the single component dynamics in the absence or presence of the well with varying depths.
    We showcase the centrality of each of the ingredients leveraged herein. 
    Numerical corroborations and a theoretical basis for our findings are provided through three-dimensional simulations emulating the experimental setting and via a one-dimensional analysis further exploring its evolution dynamics.
\end{abstract}

\maketitle

\paragraph*{\bf Introduction.} \label{sec:intro} 

The fascination with rogue or freak waves has a time-honored history that can be argued to artistically go all the way back to Hokusai's famous drawing of ``The Great Wave off Kanagawa''.
In a more quantitative form, for over half a century and since the early observations~\cite{Draper1966}, 
the term ``rogue wave'' has been used for waves of elevation several times bigger than the average sea state. 
Further, and more well-documented occurrences of rogue waves have arisen in recent years and, in particular, since the notable observation
of the so-called Draupner wave~\cite{Kharif2009}. 

Recent progress has been catalyzed by a sequence of remarkable experiments in nonlinear optics, enabling the observation of rogue waves via novel detection techniques~\cite{Solli2007} and their practical use, e.g., for supercontinuum generation~\cite{Solli2008} and continued through a sequence of detailed analysis of related waveforms~\cite{Kibler2010,Kibler2012,DeVore2013,Frisquet2016,Tikan2017}.
One candidate solution for rogue waves appearing in nature is the Peregrine soliton (PS)~\cite{Peregrine1983}.
Subsequently, both fundamental, but also more complex (higher-order) rogue-wave patterns were observed in highly controlled fluid experiments~\cite{Chabchoub2011,Chabchoub2012,Chabchoub2014}, including the very recreation of the Draupner wave~\cite{McAllister2019}.
In turn, this progress prompted related investigations in other fields, including plasmas~\cite{Bailung2011,Sabry2012,Tolba2015} 
and the associated activity has more recently been summarized in a number of related reviews~\cite{Yan2012a,Onorato2013,Dudley2014,Mihalache2017,Dudley2019,Tikan2021a}.

Bose-Einstein condensates (BECs)~\cite{Pethick2008,Pitaevskii2003}
have constituted a fertile playground where {various types of nonlinear waves}, including bright and dark solitons, vortices, vortex lines and rings, among others~\cite{Kevrekidis2015}, have been realized experimentally at a mean-field level.
Importantly, the above list also extends to numerous salient features of attractive condensates, including the formation of bright solitons~\cite{Khaykovich2002}, the modulational instability that may produce trains thereof~\cite{Strecker2002,Strecker2003,Everitt2017}, or the nature of their interactions and collisions~\cite{Nguyen2014}. 
Yet, to the best of our knowledge, the creation of one of the most quintessential nonlinear waveforms, i.e., the PS~\cite{Peregrine1983}, a structure localized in time and space that emerges from a modulationally unstable background and decays back to it, 
has remained elusive. 
This situation may be attributed to numerous key factors associated with the fairly precise control needed to produce such an entity.
Such factors include the structure's modulationally unstable background, the temporally localized nature of its existence (together with the typically destructive 
imaging), and the ``dimensionality reduction'' from three-dimensions (3D) to quasi-one-dimension (1D) and its impact on the resulting dynamics.

\begin{figure*}[t]
\centering
\includegraphics{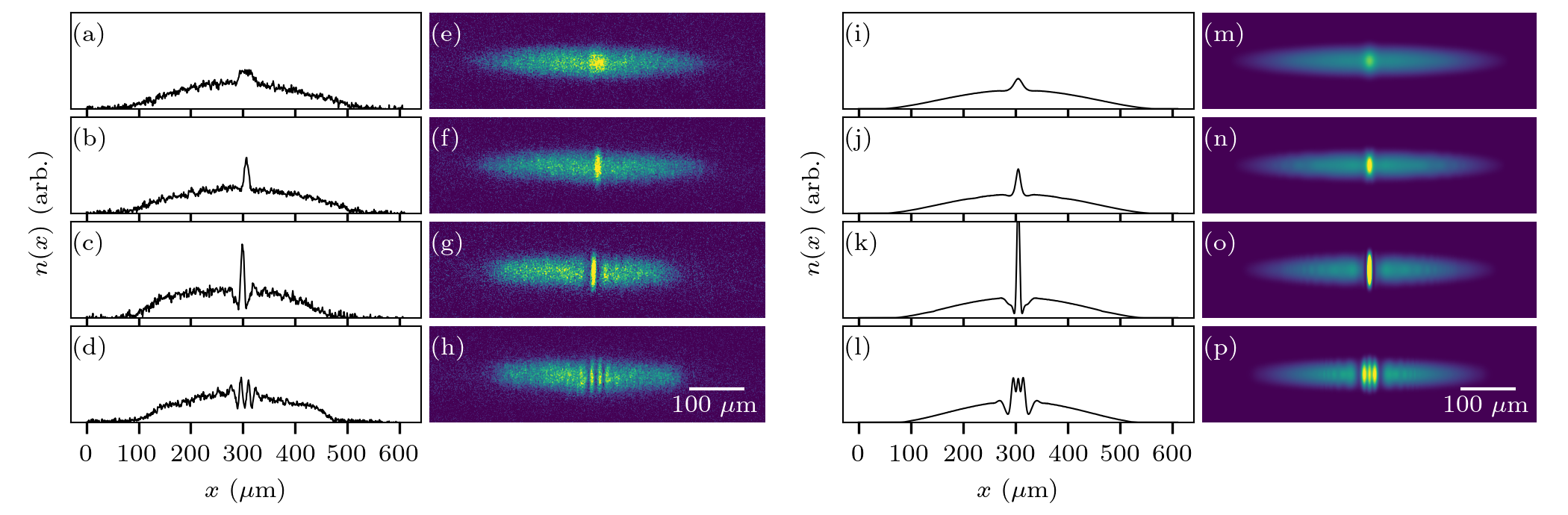} 
\caption{
    Comparison between (a)--(h) experimental and (i)--(p) numerical observations for the emergence of the PS.
    (a)--(d) Cross sections of (e)--(h) showing {single-shot} absorption images after {10, 30, 65, and 85~ms} of evolution, respectively, with an additional 9~ms of free expansion for imaging. 
    (i)--(l) Cross sections of (m)--(p) represent the density profiles obtained from the 3D mean-field  simulations under the experimental conditions.
    The vertical axis in the numerical images has been stretched for comparison with the experiment.
}
\label{fig:fig1}
\end{figure*}

The aim of the present work is to overcome these major obstacles and report the first experimental observation of the PS in BECs. 
To do so, we leverage a number of crucial ingredients.
Adapting the earlier idea of a two-component {\it self-defocusing} but immiscible setting consisting of a majority and a minority component creates an {\it effectively self-focusing medium}  for the minority component~\cite{Dutton2005,Bakkali-Hassani2021}.
This approach was utilized in two spatial dimensions to produce the well-known Townes soliton~\cite{Chiao1964} that prompted the theoretical proposal of the PS realization~\cite{Romero-Ros2022a}.

We experimentally deploy a highly elongated trap geometry with an initial (weak) potential well at the condensate center.
This well seeds the modulational instability of the minority component, providing a reproducible focal point for the spontaneous reshaping of the associated wave function into a PS, before eventually the modulationally unstable dynamics takes over and leads to the emergence of multiple peaks. 
Our numerical 3D and 1D analysis of the setting corroborates the nature of our experimental observations, while providing information about the phase structure.
Moreover, we provide experimental evidence for the centrality of each of our above-mentioned experimental ingredients, since the absence of any one of them is detrimental to the PS formation.

%--------------------------------------------------------------------------------------------------------
\paragraph*{\bf Experimental results.} 
\label{sec:exp}
We experimentally demonstrate the formation of the PS in a $^{87}$Rb BEC of  $N\approx9\times 10^5$ atoms where all interatomic interactions are repulsive. 
Initially, the atoms occupy the single hyperfine state $\lvert F, m_F\rangle = \lvert 1,-1\rangle$.
The BEC is confined in a highly elongated harmonic trap with frequencies $(\omega_x,\omega_y,\omega_z) = 2\pi\times(2.5, 245,258)$~Hz. 
The 100:1 aspect ratio of the optical trap ensures effectively 1D dynamics, leaving at most collective excitations (i.e., absence of any nonlinear structure) along the transverse direction observed in experiment and confirmed numerically. 
An additional attractive optical potential is present in the central part of the BEC producing a small density hump in the center of the cloud; See Supplemental Material (SM)~\cite{supplement} for further details. 
This optical potential, characterized by waists $s_x \approx 13$~$\rm{\mu m}$ and $s_y \approx 25$~$\rm{\mu m}$ and approximate depth of 97~nK, is radially uniform but has a Gaussian shape along the long axis of the BEC.
From this static initial condition with chemical potential $\mu \approx 97$nK~\cite{supplement}, instability is induced by rapidly transferring a small fraction ($\sim$15\%) of the atoms to the $\ket{2,0}$ state with a 55~$\mu$s microwave pulse, and transferring the remaining atoms to the $\ket{1,0}$ state in a 102~$\mu$s RF pulse.
Both pulses are applied uniformly across the whole BEC. 

In the following, we focus on the dynamics of the $\ket{2,0}$ hyperfine state (minority component) for which an effective self-focusing description applies. 
Snapshots of the corresponding density distributions 
are presented both in experiment and theory in Fig.~\ref{fig:fig1}.
The experimental images [Figs.~\ref{fig:fig1}(a)--(h)] include an additional 9~ms of time of flight to avoid image saturation of the high density peak.
The initially prepared Gaussian hump in the center of the BEC is seen to evolve into a narrow, high peak flanked by two clear dips on either side, after approximately 65~ms [Figs.~\ref{fig:fig1}(c,g)].
These dips are a characteristic feature of a PS and are related to the formation of a $\pi$ phase jump of the wave function in the peak region relative to the surrounding BEC, leading to destructive interference at the position of the dips [see also Fig.~\ref{fig:num_well_off}].
Subsequently, the peak height decreases, leading to the emergence of side peaks and excitations on either side around 85~ms [Figs.~\ref{fig:fig1}(d,h)].
We note that the observed timescales are highly reproducible, indicating that the dynamics are not triggered by a random instability, but rather are a consequence of the initial conditions prepared in the experiment.
This is also confirmed by our 3D simulations [Figs.~\ref{fig:fig1}(i)--(p)], further discussed below.

%--------------------------------------------------------------------------------------------------------

\paragraph*{\bf Mean-field dynamics.} 
\label{sec:model}

Following the experimental conditions, we consider a $^{87}$Rb BEC in the aforementioned hyperfine states with a spin population imbalance of 85\%-15\%. 
To model the dynamical generation of the PS, we employ two coupled 3D Gross-Pitaevskii equations~\cite{Pitaevskii2003,Pethick2008,Kevrekidis2015},
\vspace*{-1ex}
\begin{multline}
    \label{eq:CGPE}
    i\hbar\partial_t\Psi_F(\textbf{r},t) = \bigg[-\frac{\hbar^2}{2m} \nabla_{\textbf{r}}^2 + V(\textbf{r})+V_G(\textbf{r}) 
    \\
      + \sum_{F'=1}^2 g_{FF'}|\Psi_{F'}(\textbf{r},t)|^2 \bigg] \Psi_F(\textbf{r},t)\,.\qquad
\end{multline}
Here, $\Psi_F(\textbf{r},t)$ is the 3D mean-field wave function with $F=1,2$ denoting each hyperfine state, $\textbf{r}=(x,y,z)$, and $m$ is the atomic mass. 
The external trap is given by $V(\textbf{r})=\sum_{\alpha=x,y,z} m\omega_\alpha^2\alpha^2/2$, and the coupling constants $g_{FF'}=4 \pi N_{F'} \hbar^2 a_{FF'}/m$ refer to the intra- $(F=F')$ and interspecies ($F\neq F'$) interaction strengths, with $a_{FF'}$ being the 3D $s$-wave scattering lengths, and $N_F$ is the atom number in the $F$ spin channel. 
Specifically, the scattering lengths corresponding to the experimental setup are $a_{11}=100.86a_0$, $a_{22}=94.57a_0$, and $a_{12}=a_{21}=98.9a_0$, where $a_0$ designates the Bohr radius. 
These coefficients give rise to an {\it effective} attractive nonlinear  coefficient $a_\mathrm{eff}=a_{22}-a_{12}^2/a_{11}<0$, allowing for a reduced single-component description of the minority component~\cite{Dutton2005,Bakkali-Hassani2021}.

Consequently, our system now supports the emergence of focusing nonlinear phenomena such as the PS. 
%for which, 
Neglecting the transverse coordinate dependence, the form of the PS is given by~\cite{Peregrine1983}
%the minority component is approximately given by the form
%
\begin{equation}
    \Psi_P(x,t) = \sqrt{P_0}
    \qty[1 - \frac{4 \qty(1 + 2 i \frac{t-t_0}{T_P})}{1+4 \qty(\frac{x-x_0}{L_P})^2 + 4\qty(\frac{t-t_0}{T_P})^2}] e^{i \frac{t-t_0}{T_P}},
    \label{eq:peregrine}
\end{equation} 
where $T_P/\hbar=L_P^2m/\hbar^2=1/(|g_{{\rm eff}}| P_0)$. Here, $T_P$ and $L_P$ are the characteristic scales of time and space of the PS solution, respectively. 
$P_0$ represents the background density of the minority component in a homogeneous system, and $g_{{\rm eff}}=g_{22}^{(1{\rm D})}-\qty(g_{12}^{(1{\rm D})})^2/g_{11}^{(1{\rm D})}$ denotes the effective 1D interaction in the single-component description, see also SM~\cite{supplement}.

To dynamically seed the PS nucleation, we employ the optically induced Gaussian well 
$V_G(\textbf{r})=-V_0 \exp{-2\left[\left(x/s_x\right)^2+\left(y/s_y\right)^2\right]}$.
%$V_G(\textbf{r})=-V_0 \exp{-2\left[\left(\frac{x}{s_x}\right)^2+\left(\frac{y}{s_y}\right)^2\right]}$.
%
The widths and the potential depth $V_0$  are fixed in accordance with the experimental setup. 
%while the potential depth $V_0$ is varied to dynamically control the emergence of 
%the PS at earlier evolution times. 
Note that the transverse spatial profile
of the Gaussian potential does not significantly affect
the PS generation, in line with the experimental observations, as long as its width is larger than the transverse spatial extent of the
BEC.

We initially place all $N$ atoms in the $\ket{1,-1}$ state and identify the ground state of this system in the presence of the optical well utilizing the time-independent version of Eq.~(\ref{eq:CGPE}). 
We then instantaneously transfer a fraction of typically $15 \%$ ($85 \%$) atoms to the $\ket{2,0}$ ($\ket{1,0}$) state, thus emulating the RF experimental process. 
Additionally, we approximately account for the experimental thermal fraction ($<10 \%$) and for the observed atom-loss rate in $\ket{2,0}$ of around $0.23\%$ per ms (see  SM~\cite{supplement}).
The two-component system is then allowed to evolve according to Eq.~(\ref{eq:CGPE}). 
Initially, the dynamical evolution of the stationary states described above entails the counterpropagating emission of sound waves [cf.\ Figs.\ref{fig:fig1}(i,m) and \ref{fig:fig1}(j,n)] with the subsequent PS generation reaching maximal amplitude around {$t\approx 70\,\rm{ms}$} [see the slightly earlier snapshots in Figs.\ref{fig:fig1}(k,o)], before its structural deformation toward three equidistant peaks [Figs.\ref{fig:fig1}(l,p)]. 
{A clear agreement with the experimental PS realization and the overall dynamics [Figs.~\ref{fig:fig1} (a)--(h)] is observed.
Any residual deviations in the intensity of the PS are principally traced back to the time of flight performed in the experiment but not taken into account in the simulations.}

%--------------------------------------------------------------------------------------------------------
\paragraph*{\bf Controllability of Peregrine generation.}  
\label{sec:emergence}

\begin{figure}[t]
\centering
\includegraphics[width=\linewidth]{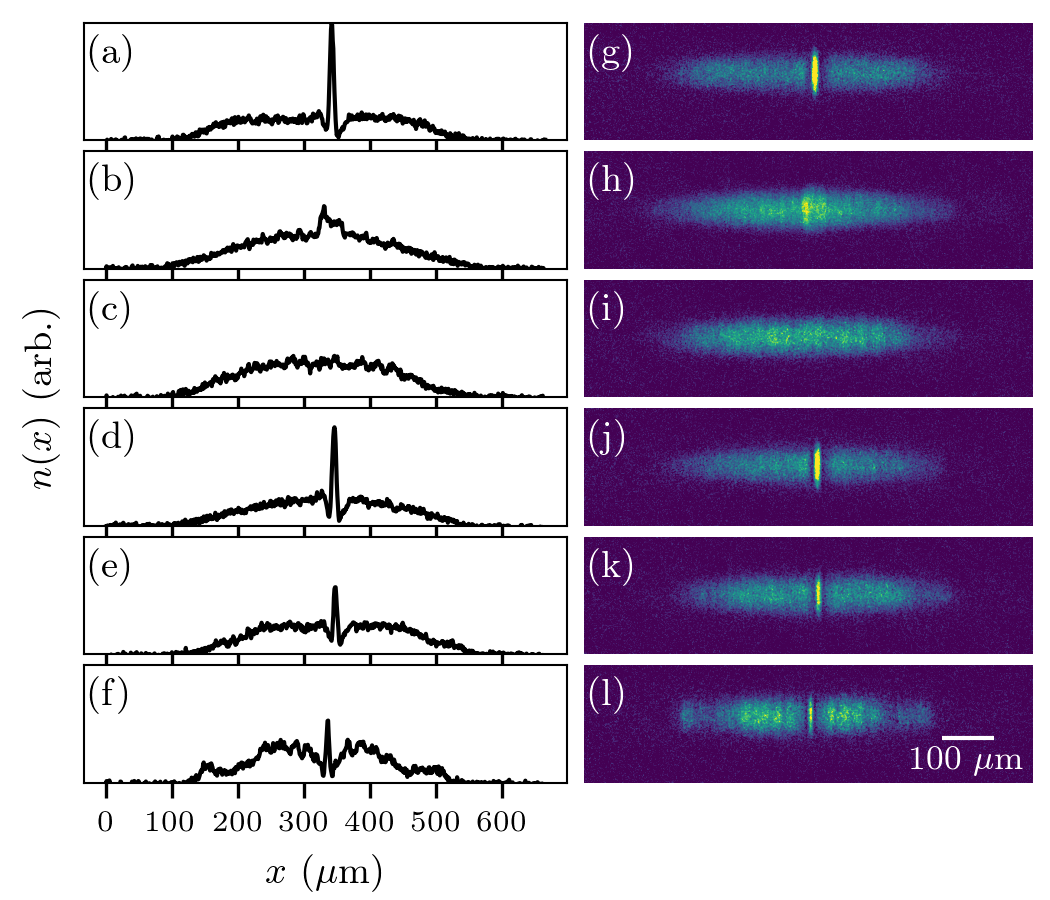}
\caption{
    Impact of the optical trap features on PS nucleation.
    (a), (g)~Standard PS sequence after 50~ms of evolution.
    (b), (h)~All atoms in a single component ($\ket{1,0}$) showing no PS formation.
    (c), (i)~Minority component prepared without the potential well leading to the absence of PS. 
    (d), (j) A PS forming in the $\ket{1,-1}\&\ket{2,0}$ mixture after 50~ms of evolution (instead of the $\ket{1,0}\&\ket{2,0}$ mixture).
    (e), (k) Well depth cut by half compared to panel~(a), then 80~ms evolution. 
    (f), (l) Well [with the same depth as in panel~(a)] switched off at 20~ms, with the image taken after 110~ms (i.e., 90~ms after the well was switched off). 
    For further details on the interplay of PS generation and the well characteristics see SM~\cite{supplement}.
}
\label{fig:fig2}
\end{figure}

To unveil the necessary conditions for the formation of a PS, Fig.~\ref{fig:fig2} presents a composition of various alterations of the experimental procedure discussed above.
As a baseline for comparison, Figs.~\ref{fig:fig2}(a,g) show a PS beginning to form under the conditions described in Fig.~\ref{fig:fig1} after 50~ms of evolution.
If an identical experiment is performed but with a single-component cloud, no PS is observed [Figs.~\ref{fig:fig2}(b,h)], demonstrating the key role of interspecies interactions for the emergent dynamics. 
The deformation of the initially Gaussian bulge is due to expansion during time of flight. 
Specifically, the initial Gaussian shaped density hump spreads out, leading to sound wave pulses propagating away from each other. 
Also, when conducting experiments with the two-component mixture in the absence of the well, instability takes longer to set in and no PS forms within accessible timescales [Figs.~\ref{fig:fig2}(c,i)].

Having identified the presence of the optical well and the genuine two-component mixture as key ingredients, we can further elucidate their roles.
Figures~\ref{fig:fig2}(d,j)
show a mixture of 15\% of atoms in the $\ket{2,0}$ state embedded in a 85\% background of atoms in the $\ket{1,-1}$ state [as opposed to $\ket{2,0}$ and $\ket{1,0}$ atoms used for Figs.~\ref{fig:fig2}(a,g)].
The dynamical generation of the PS is again clearly observed, although this mixture is characterized by a less attractive effective scattering length of $a_\textrm{eff} = -1.34a_0$ for the $\ket{2,0}$ atoms, as compared to $-2.41 a_0$ for the
$\ket{2,0}$ atoms embedded in a $\ket{1,0}$ background. 
The formation of a PS, as discussed above, is not highly specific to some of the exact parameters of the Gaussian well, e.g., if the well depth is reduced by a factor of two, the PS still emerges, but at later evolution times.
In particular, in Figs.~\ref{fig:fig2}(e,k) the PS starts to manifest after 80~ms of evolution time, compared to the approximately 50~ms needed in the case depicted in Figs.~\ref{fig:fig2}(a,g).

Importantly, the PS can emerge even if the well is only present for a short time after the initial preparation of the mixture, 
and it is then switched off.
Figures~\ref{fig:fig2}(f,l) showcase a pertinent example, where the well was switched off abruptly at 20~ms after the preparation of the atomic mixture, and the image was taken after an additional evolution time of 90~ms after the switch-off (see also the discussion in SM~\cite{supplement}).  
This comparison demonstrates that the continued presence of the potential well is not required: 
the well only serves to ``seed'' the relevant dynamics leading to the PS generation. 
The possibility to trigger the dynamics in a controlled way is a powerful feature of our experimental setting, which enables us to produce the PS in a highly repeatable way, making it possible to study its time evolution.
The instrumental role played by the well is further elucidated through more elaborated numerical investigations of the impact of its characteristics provided in Fig.~\ref{fig:num_well_off}.
%

%--------------------------------------------------------------------------------------------------------
\paragraph*{\bf Further characterization of the Peregrine.}
\begin{figure}[t]
\centering
\includegraphics[width=\linewidth]{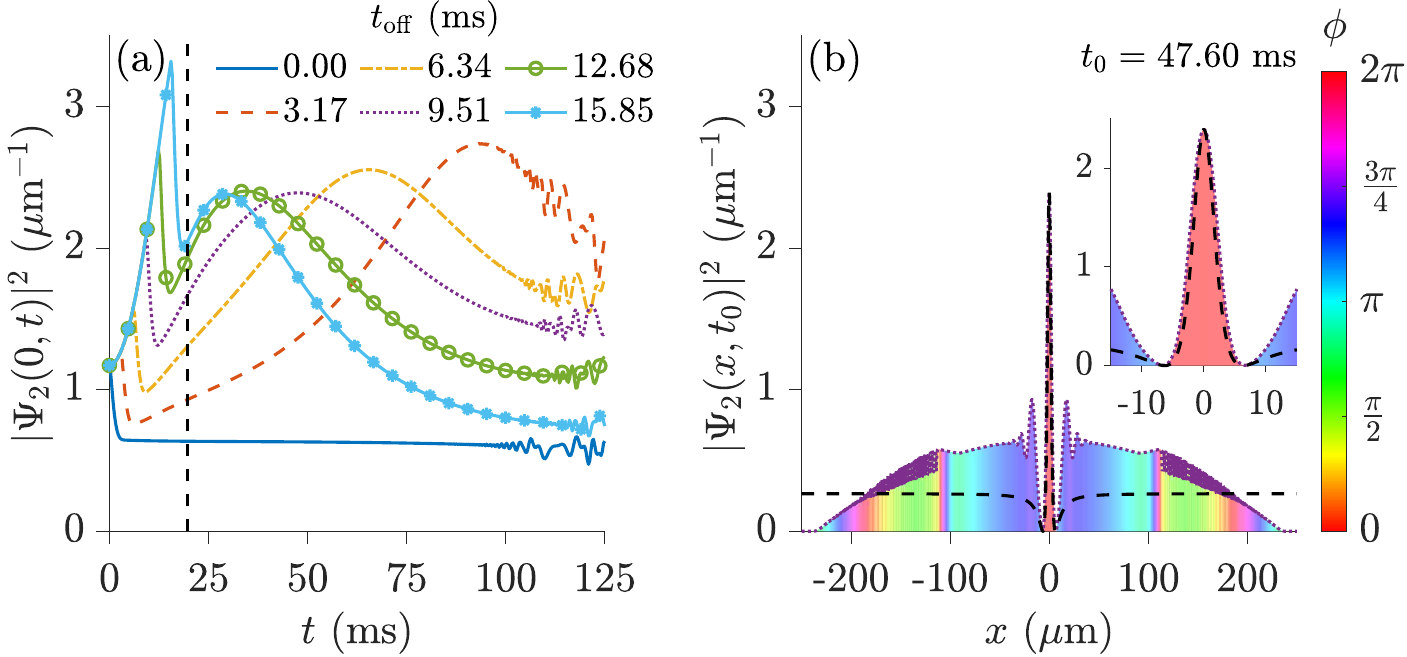}
\caption{
    1D simulations of the minority component  switching off the Gaussian well at indicated times before the expected nucleation of the PS (vertical black dashed line). 
    (a) Time evolution of the central density, $|\Psi_2|^2$, of $\ket{2,0}$. 
    (b) Snapshot of $|\Psi_2|^2$ at the time instant of the PS formation after the well switch-off at $t=9.51$~ms. 
    The color gradient denotes the phase of $\Psi_2$.     
    A magnification of the central region in the inset, showcases the good agreement between $|\Psi_2|^2$ and the analytical PS solution~\eqref{eq:peregrine} (black dashed line) and the characteristic $\pi$ phase jump between the core and the wings of the PS.
    The Gaussian well parameters used here are $V_0=60$~nK and $s_x=4.8$~$\mu$m.
}
\label{fig:num_well_off}
\end{figure}

Leveraging the 1D nature of the PS, we additionally employ a 1D reduction of Eq.~\eqref{eq:CGPE} to further numerically characterize the features of the PS in the context of these experiments.
Here, we follow the experimental procedure described above while averaging over the transverse coordinates (see SM for details~\cite{supplement}).

Figure~\ref{fig:num_well_off} demonstrates how the presence of the well assists in controllably seeding the emergence of the 
PS.
In this particular case, we employ a well with $V_0=60$~nK
and $s_x=4.8$~$\mu$m.
In Fig.~\ref{fig:num_well_off}(a) we present the time evolution of the central density of the minority component, $|\Psi_2(0,t)|^2$, when switching off the well at various time instants, $t_\textrm{off}$ (see legend), 
before the PS nucleation would occur if the well was always present (vertical dashed line at $t=19.68$~ms).
In all cases, the PS emerges at some time, $t_0$, after switching off the well.
The exception is $t_{\textrm{off}}=0$ms (blue solid line), for which no PS forms.
This supports the fact that, without the well, the above initial condition is not sufficient to form the PS.
The earlier the switch-off, the later the PS emerges.
Note that the process of switching off the well generates shock waves and their effect is visible after $t>100$~ms.

To better understand the structure and properties of the PS, in Fig.~\ref{fig:num_well_off}(b), we provide an instantaneous density profile of $\ket{2,0}$ and its corresponding phase (depicted by a color gradient) at $t_0$ when switching off the well at $t_\textrm{off}=9.51$~ms.
Additionally, we provide the profile of Eq.~\eqref{eq:peregrine} with $P_0=\max(|\Psi_2(x,t_0)|^2)/9$ (black dashed lines) to compare the emerging structure with the analytical PS solution.  
A close inspection of the central region of the condensate [inset of Fig.~\ref{fig:num_well_off}(b)] evinces the excellent agreement of the PS core among the two and the telltale $\pi$ phase jump between the core and the wings of the waveform.

%--------------------------------------------------------------------------------------------------------

\paragraph*{\bf Conclusions.} 
We have experimentally demonstrated the dynamical formation of a PS in a two-component BEC featuring a suitable mixture of repulsive interactions that emulate an effective attractive environment. 
This work shows how self-focusing interactions together with an attractive well as an effective catalyst cause a time-dependent localization to emerge from a modulationally unstable background resulting in the realization of a PS.
Utilizing the attractive potential well it was possible to reproducibly and rapidly, i.e., comfortably within the condensate lifetimes, produce such wave structures in a highly controlled manner. 
A single repulsive component, not being modulationally unstable, is unable to produce such a phenomenon. 
Importantly, our experimental observations are in good quantitative agreement with 3D mean-field simulations. 
Simultaneously, a systematic 1D analysis revealed additional features of the phenomenology, 
such as the telltale phase gradient across the PS, and a detailed examination of the effect of switching off the well at different times. 

Our platform paves the way for a closer inspection of rogue waves and higher-order rogue structures~\cite{Chabchoub2012},
or rogue waves in other ultracold atomic gas implementations such as intrinsically attractive 
BECs~\cite{Khaykovich2002,Strecker2003,Everitt2017,Nguyen2014}. 
A natural question is the persistence of the PS generation in the dimensional crossover to 3D and how (parametrically) the 2D or 3D character comes into play. 
Another direction would be to extend these considerations to a larger number of components (e.g., spinor condensates~\cite{Kawaguchi2012,Stamper-Kurn2013}), to reveal the interplay of magnetic excitations and possibly emergent spin domains on the PS formation.
Yet another possibility may be to study the formation of the mixed-bubble phase~\cite{Naidon2021,mistakidis2023few,Sturmer2022} that is inherently related to the presence of quantum fluctuations and occurs at the immiscibility threshold.  

%--------------------------------------------------------------------------------------------------------
\paragraph*{\bf Acknowledgements.}
We acknowledge fruitful discussions lending context to this work at the Dispersive Hydrodynamics Program (2022) hosted by the Isaac Newton Institute for Mathematical Sciences. 
This material is based upon work supported by the U.S.\ National Science Foundation (NSF) under the awards  PHY-2110030 and DMS-2204702 (P.G.K.). 
S.I.M. acknowledges support from the NSF through a grant for ITAMP at Harvard University.
P.E. acknowledges support from the NSF through Grant No. PHY-1912540 and from the
Ralph G. Yount Distinguished Professorship at WSU.
G.B. acknowledges support from the NSF through Grant No. DMS-2004987.
The work of P.S. was funded by the Deutsche Forschungsgemeinschaft (German Research Foundation) under grant SFB-925 – project 170620586. 

%--------------------------------------------------------------------------------------------------------
\bibliographystyle{apsrev4-1}
\bibliography{Peregrine_waves.bib}

%%%%%%%%%% Merge with supplemental materials %%%%%%%%%%
\onecolumngrid
\newpage
\clearpage
\begin{center}
	\large{\bf{Supplemental Material}}
\end{center}
\twocolumngrid
%%%%%%%%%% Merge with supplemental materials %%%%%%%%%%
%%%%%%%%%% Prefix a "S" to all equations, figures, tables and reset the counter %%%%%%%%%%
\setcounter{equation}{0}
\setcounter{figure}{0}
\setcounter{table}{0}
\setcounter{page}{1}
\makeatletter
\renewcommand{\theequation}{S\arabic{equation}}
\renewcommand{\thefigure}{S\arabic{figure}}
%%%%%%%%%% Prefix a "S" to all equations, figures, tables and reset the counter %%%%%%%%%%

\section{Experimental methods and supplementary results}\label{sec:sup1}
\begin{figure}[t]
	\centering
	\includegraphics[width=\linewidth]{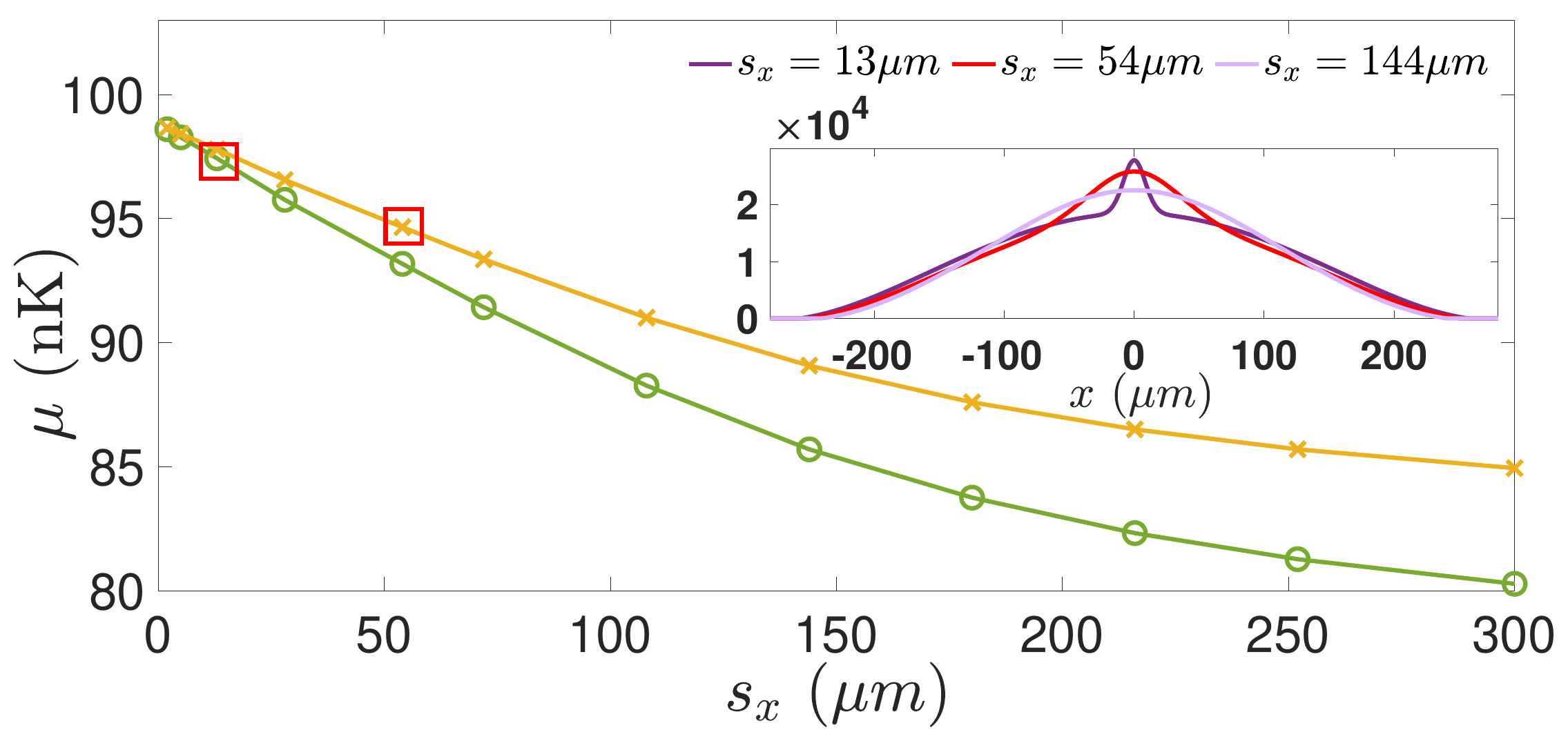}
	\caption{{Chemical potential variation of the initial state as a function of the width, $s_x$, of the attractive Gaussian potential along the long axis of the BEC. 
			Distinct solid lines correspond to the  specific choices of $V_0$ (see legend) employed in the main text. 
			Red rectangles mark the values used in Figs.~1 and 4 of the main text. 
			Notice that $s_x=300$~$\mu$m $\approx R_{TF}$ representing the Thomas-Fermi radius in the absence of the Gaussian well. 
			The inset depicts characteristic integrated density profiles of the initial state for different $s_x$ (see legend). 
			Other parameters used are the same as in Fig.1 of the main text.}}
	\label{fig:chem_pot_variation}
\end{figure}

Our experiments are conducted with $^{87}$Rb atoms in the $F=1$ and $F=2$ spin states. Initially, BECs containing approximately $N=9 \times 10^5$ atoms in the $\ket{F, m_F}=\ket{1,-1}$ hyperfine state are confined in a highly elongated optical dipole trap with harmonic trap frequencies of $(\omega_x,\omega_y,\omega_z) = 2\pi\times(2.5, 245,258)$~Hz. 
A magnetic bias field of 10~G is applied in the vertical direction. During the evaporative cooling that leads to the formation of the BEC, an attractive Gaussian potential is continuously superimposed on the optical dipole trap.
This potential is generated by an elliptical dipole beam with a wavelength of 850~nm and is characterized by a Gaussian width of {$s_x=13~\mu$m} along the long axis of the BEC and {$s_y=25~\mu$m} along the tightly confined horizontal direction.
The beam propagates in the vertical direction. Therefore, in both tightly confined directions the beam is significantly larger than the diameter of the BEC (approximately 3~$\mu$m), and its profile is relevant only along the weakly confined, long axis of the BEC. The potential depth produced by this dipole beam is {$V_0=40$}~nK (for Fig.~1 and 2 in the main text with the exception of Fig.~2(e,k) where the potential depth was reduced by a factor of two), and {$V_0=30$}~nK for the case of the wide well in Fig.~\ref{fig:fig4}. 
This potential produces a density hump in the center of the BEC. This leads to a static initial condition with no detectable dynamics{, see in particular the inset of Fig.~\ref{fig:chem_pot_variation}. Notice that larger widths lead to  less deformed yet broader initial state configurations. 
	The values of the respective chemical potential, $\mu$, of this initial state are provided in Fig.~\ref{fig:chem_pot_variation} for different widths, $s_x$, of the attractive Gaussian potential. 
	All other parameters are kept fixed. 
	As it can be seen, $\mu$ is reduced for increasing $s_x$ since the latter enforces a wider Thomas-Fermi profile. 
	On the other hand, $\mu$ is shifted to  larger values for smaller $V_0$ and fixed $s_x$.} 

\begin{figure*}[t]
	\centering
	\includegraphics[width=\linewidth]{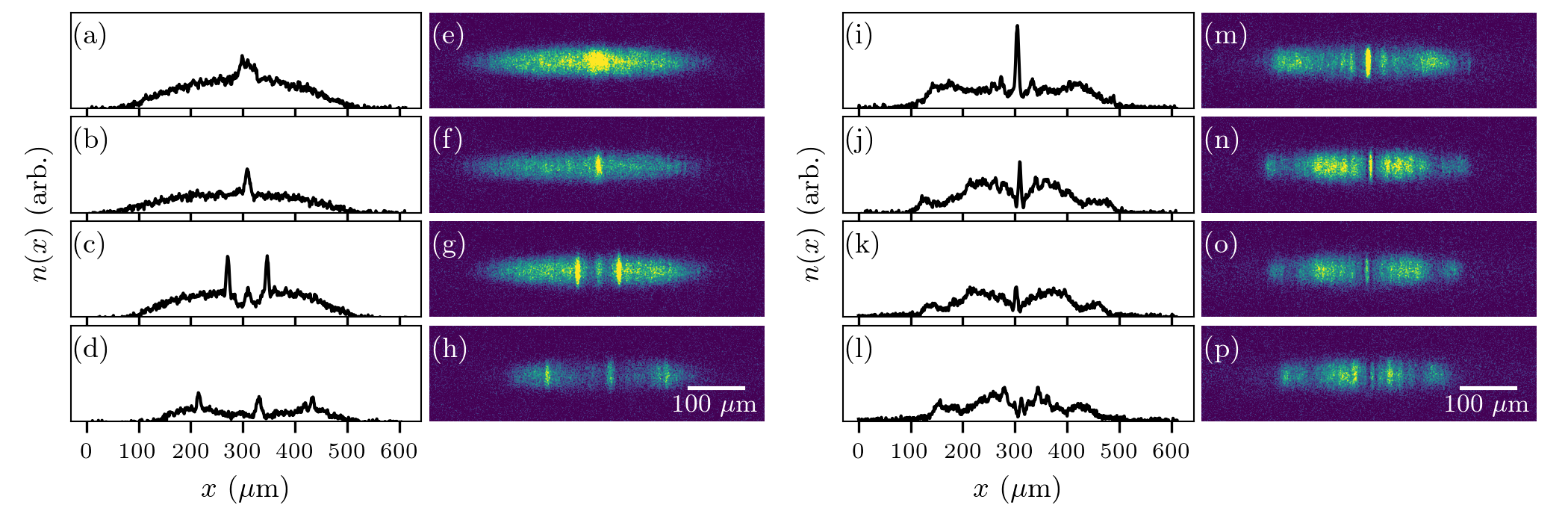}
	\caption{
		Time evolution for the case of an early switch-off of the well. The BEC is prepared with the same parameters as in Fig.~1 of the main text. The different panels show images taken after (a, e) 10~ms, (b, f) 20~ms, (c, g) 30~ms, (d, h) 60~ms, (i, m) 100~ms, (j, n) 110~ms, (k, o) 130~ms, and (l, p) 150~ms. The well is switched off after 20~ms, corresponding to panels (b, f). The images depict the atoms in the  $\ket{F, m_F}=\ket{2,0}$ state and are taken with 9~ms time of flight.
	}
	\label{fig:well_off}
\end{figure*}

A rapid microwave and RF pulse sequence then generates a two-component mixture comprised of  85\% of all atoms in the $\ket{F, m_F}=\ket{1,0}$ state  and 15\% of all atoms in the $\ket{F, m_F}=\ket{2,0}$. {We remark that the mean-field interaction coefficients, $g_{ij}=4 \pi \hbar^2 a_{ij}/m_\qty(^{87}{\rm Rb})$, correspond to  $g_{11}=5.16 \times 10^{-51}~({\rm kg~  m^5/s^2})$, $g_{22}= 4.84 \times 10^{-51}~({\rm kg~m^5/s^2})$ and $g_{12}=g_{21}=5.06 \times 10^{-51}~({\rm kg~m^5/s^2})$.} {State purity has been confirmed using Stern Gerlach imaging.} This initiates the dynamics leading to the formation of a Peregrine soliton.
As described in the main text, the attractive Gaussian potential in the center of the BEC only serves to induce the initial dynamics.
It can either be left on, or it can be switched off after a brief initial period, even before a clear Peregrine peak has formed (see the example shown in Fig.~\ref{fig:well_off}). 
In both cases the formation of a Peregrine soliton can be observed, though after different evolution times. 

In the example of Fig.~\ref{fig:well_off}, which will be described in greater detail later in this Supplement, the well was switched off 20~ms after the formation of the mixture, which is clearly before a Peregrine soliton would reach its peak height in the continued presence of the well. As a consequence of the switch-off, sound excitations are seen to propagate outwards to the edge of the BEC. Their propagation speed provides a sense for mean-field time scales of the system. The central peak still continues to grow after the switch-off, reaching a peak height approximately 80~ms after the switch-off in this example.   
Also, the observation of the Peregrine soliton is not specific to the chosen hyperfine mixture [a mixture of atoms in the $\ket{F, m_F}=\ket{1,-1}$ and $\ket{2,0}$ has also been used successfully, see Fig.~2(d,j) of the main text] and also does not critically depend on the specific component ratios.
The slight asymmetry observed in the wings of the Peregrine solitons, as seen e.g. in Fig.~1(c,g) of the main text, can possibly be attributed to a minute and otherwise undetectable amount of counterflow between the two components of the mixture.

\begin{figure}[t]
	\centering
	\includegraphics{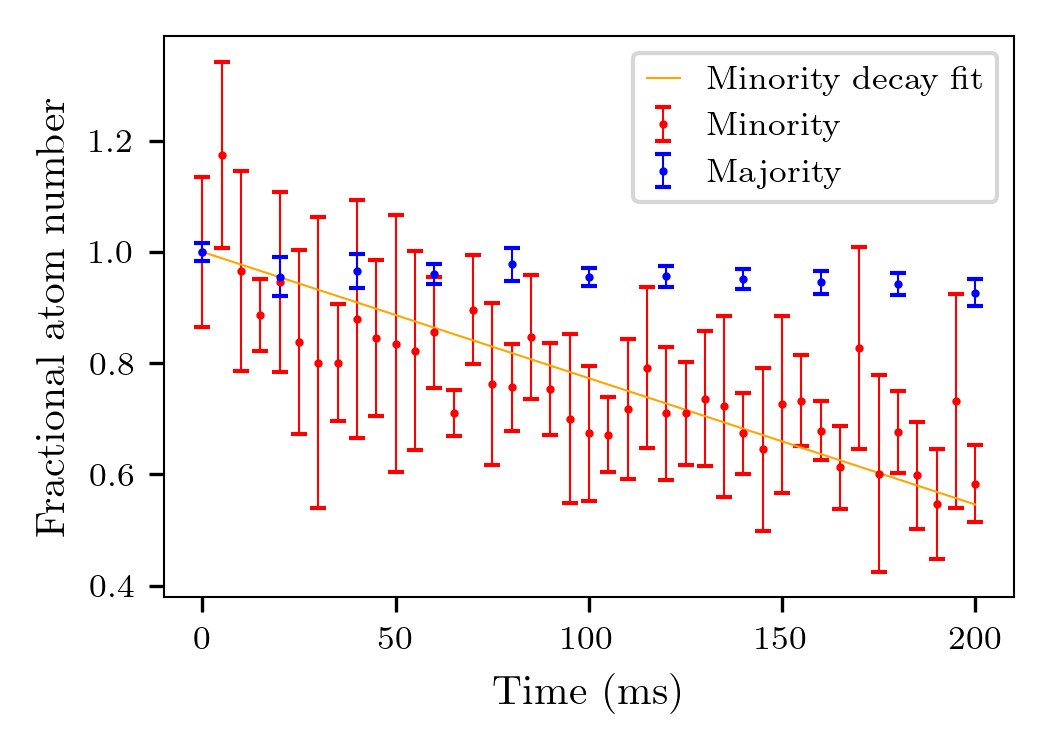}
	\caption{{Fractional atom number in the majority and minority components (see legend) in the case where the attractive well is left on throughout the dynamics. The majority component shows negligible atom loss over this time scale, while the minority component features nearly linear decrease at a rate of 0.23\% per ms. Error bars are given by one standard deviation over five independent realizations.}}
	\label{fig:atomlossrate}
\end{figure}

{The fractional atom number for both the majority and minority components is shown in Fig.~\ref{fig:atomlossrate} in the case where the attractive barrier is left on throughout the experiment.
	The majority component atom number shows minor  changes, comparable to a single component BEC held in a similar trap. 
	{The minority component experiences an atom number decay rate of approximately 0.23\% per ms, resulting in roughly half of the minority population remaining in the system after 200~ms.}
	To shed light on the impact of these loss rates, we take them into account in the corresponding 3D mean-field simulations through the appropriate renormalization of the corresponding component wave function every millisecond such that they follow the fractional atom loss rates provided in Fig.~\ref{fig:atomlossrate}. 
	It turns out that there is only a minor effect on the timescale of Peregrine formation (delayed by $\sim 2$~ms compared to the zero loss case), but it drastically affects the amplitude of the ensuing wave which is found, as expected, to be significantly reduced.}

Lastly, our experiments with a “too wide” attractive well show modulationally unstable dynamics which, including thermal effects, do not appear to manifest a clearly distinguishable isolated Peregrine solitary wave within the time scales of experimental observation. The Gaussian widths resulting in the corresponding experimental observations shown in Figs.~\ref{fig:fig4} are $s_x \approx 54$~$\mu$m and $s_y \approx 25$~$\mu$m. 
The initial wide pattern [Figs.~\ref{fig:fig4} (a,e)] having $\mu \approx 94$~nK features in the course of the evolution a decrease in its intensity with the central peak becoming more transparent around 90~ms [Figs.~\ref{fig:fig4}(b,f)]. 
Subsequently, this central peak structure becomes gradually more prominent but instead of focusing towards a Peregrine wave signature starts to break into multiple peaks [Figs.~\ref{fig:fig4}(c,g)] which later on interfere [Figs.~\ref{fig:fig4}(d,h)]. As was shown in~\cite{Bertola2013,Grimshaw2013}, the significance of the Peregrine soliton is that it quantitatively captures the generic mechanism of spontaneous self-focusing and gradient catastrophe formation in attractive nonlinear media. The dynamics described above is also a precursor of a nonlinear modulational instability behavior similarly to Ref.~\cite{Biondini2016b}, whose early signature is also visible in the later stages of the time evolution, as seen in Figs. 1(d,h) and 1(l,p).

\begin{figure}[t]
	\centering
	\includegraphics{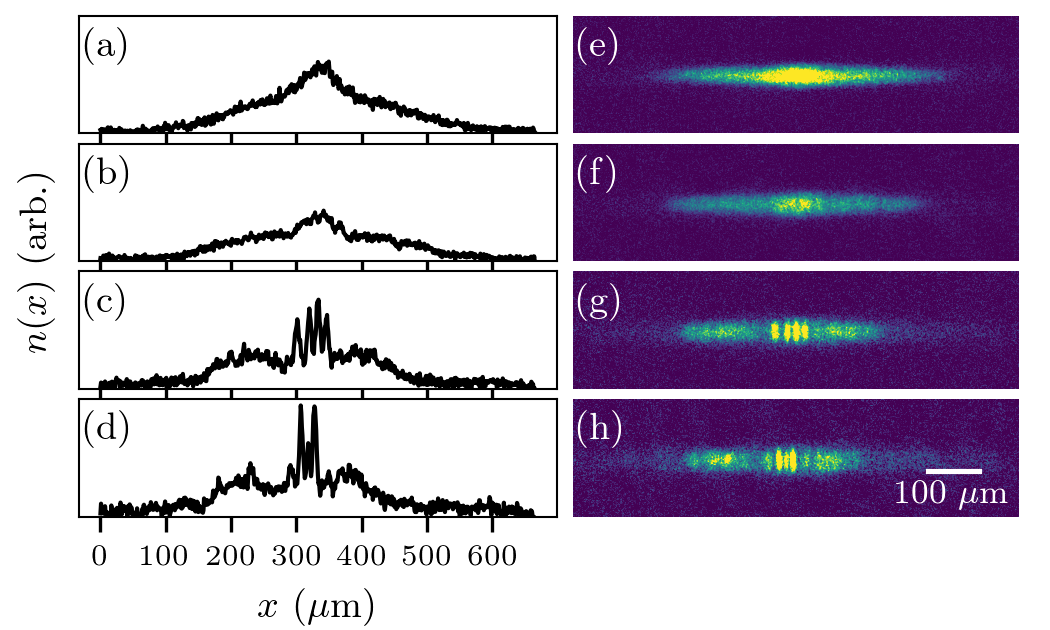}
	\caption{
		Experimental observation of nucleation of solitonic features under a wider attractive well.
		(a)--(d) Cross sections of (e)--(h) showing single-shot absorption images after 0, 90, 158, and 230ms of evolution, respectively.
	}
	\label{fig:fig4}
\end{figure}

\section{Delineating the impact of different protocols} \label{sec:deviations}

{To explore the origin of deviations observed between the 3D mean-field simulations and the experiment we further checked possible experimental imperfections and mechanisms that are not emulated by the theory. 
	As such, we identified three main sources that may lead to discrepancies among theory and experiment, see also Table~\ref{Table_times}. 
	These refer to a residual thermal fraction (less than $10\%$) detected in the experiment, the associated fitting procedure to infer the Peregrine formation time extracted from the experimental images and the time of flight imaging technique that is not modeled in the simulations. Below, we explain the impact of these distinct processes on the Peregrine nucleation, while following in all cases the dynamical protocol outlined in the main text. 
	Peregrine manifestation is estimated by identifying the maximum peak density of the minority component during evolution.}

\begin{center}
	\begin{table} [h!]
		\centering
		\begin{tabular}{ |c|c|c|c| } 
			\hline
			Switch-off time & Experiment & Simulations \\
			\hline
			{$\infty$} & {$65\pm14$ ms} & {$70$ ms} \\ 
			\hline
			{$30$ ms} & {$107\pm29$ ms} & {$114$ ms} \\ 
			\hline
			{$20$ ms} & {$155\pm47$ ms} & {$178$ ms} \\ 
			\hline
			{$10$ ms} & {$>200$ ms} & {$240$ ms} \\ 
			\hline
		\end{tabular}
		\caption{{time instant of the Peregrine soliton's maximum amplitude upon switching-off at different times (first column) the Gaussian potential. 
				The time at which the Peregrine amplitude maximizes in the experiment (second column) and within the 3D mean-field simulations (third column) emulating the presence of a finite thermal fraction and atom loss of the minority component $\sim0.23$ per ms. The experimental uncertainties are given by the width of a Gaussian fit to the Peregrine's amplitude as it emerges and recedes.}}
		\label{Table_times}
	\end{table}
\end{center}

{To mimic the experimentally observed deformations of the condensate configuration due to thermal effects we exploit a weak perturbation modeled by a random noise term. 
	The latter is added to the mean-field zero temperature ground state of the F-component, $\Psi_F^G(x,y,z)$, according to the following ansatz $\Psi_F(x,y,z,t)=\Psi_F^G(x,y,z)+ \epsilon \delta(x,y,z)$. Here, $\delta (x,y,z)$ refers to a Gaussian random distribution with zero mean and unit variance, while $\epsilon \sim 0.1$ accounts for a $10 \%$ thermal fraction. 
	To adequately estimate the effect of the random distribution on the dynamics we consider a sample of at least twenty different realizations ensuring that the results do not alter upon further sampling. 
	It is found that the presence of the above thermal fraction accelerates the Peregrine formation which in the case depicted in Fig.~1 of the main text occurs at {70~ms}, while the experiment identifies the maximum Peregrine amplitude at 65~ms, see Table~\ref{Table_times}.   
	However, the intensity of the Peregrine appears to be enhanced in the 3D simulations as compared to the experimental observations. 
	Note in passing that an interesting possibility for future endeavors would be to rely on the stochastic Gross-Pitaevskii model describing  the thermal component and its interplay with the condensed fraction~\cite{Proukakis2008}.}   

{Turning to the discrepancy in terms of the Peregrine intensity we additionally explored the effect of quenching the trap frequency to lower values studying this way the consequences stemming from the time of flight imaging. 
	The aforementioned ansatz of the perturbed wave function is maintained.  
	Interestingly, the major outcomes of this process are i) to lower the intensity of the Peregrine, ii) slightly shift its occurrence to earlier times and iii) sharpening of the observed features such as the counterpropagating sound-wave emission.   
	Specifically, we establish that a trap quench to half of the initial frequency is able to quantitatively match the intensity of the Peregrine as captured by the experiment. 
	Concluding, very good agreement in terms of both occurrence times and intensity of the Peregrine requires both the presence of thermal fraction and the mimicking of the time of flight process. 
	Despite the fact that the trap quench can be used to infer a semi-quantitative understanding  of the effect of 
	the expansion it is not equivalent to the free expansion process in the experiment. 
	For this reason, the times of Peregrine formation presented in Table~\ref{Table_times} stem solely from the inclusion of the random noise term.} 

{Having the above-described knowledge, we subsequently simulated the Peregrine realization utilizing different switch-off times of the external Gaussian well. 
	Once more, good quantitative agreement for the time where the Peregrine maximizes its amplitude is identified in Table~\ref{Table_times}. }  
{It should be noted that the observed finite background in all experimental cross sections, see e.g. Fig.~\ref{fig:well_off}, of the minority component renders the measurement of the peak intensity and the time of the Peregrine formation unclear.
	The times of peak Peregrine formation for the experiment shown in Table~\ref{Table_times} are extracted by analyzing the time sequence of Peregrine formation for each initial condition.
	The Peregrine amplitude is approximated by considering the central region of the condensate's integrated cross section, around the forming Peregrine, then using the difference between the maximum and minimum densities as a proxy for the Peregrine amplitude.
	This approach captures both the extent of the soliton peak and the depth of the adjacent density depressions.
	The time of maximum amplitude was extracted by fitting a Gaussian to the Peregrine amplitude as the soliton emerges and recedes. 
	The time of maximum Peregrine amplitude then corresponds to the central point of the amplitude fits and the uncertainties given in Table~\ref{Table_times} correspond to the Gaussian width of the Peregrine formation curve in time.}

%--------------------------------------------------------------------------------------------------------

\section{Characteristics of the Peregrine within a 1D mean-field description} \label{sec:1D}

\subsection{Setup and transfer protocol} \label{sec:1D_model}

The quasi-one-dimensional (1D) geometry of the experimental setup, and its corresponding dynamics 
reported by both the experiments and the three-dimensional (3D) simulations (see main text)
allow for a 1D description of the system under consideration. 
{This can be traced back to the fact that in the transverse directions the cloud exhibits a collective motion and any kind of  nonlinear excitations is suppressed. As such, a pure 1D treatment is adequate to capture the Peregrine nucleation but not its quantitative features. For instance, it predicts earlier times of formation and a reduced Peregrine intensity due to the absence of the transverse directions. 
	Nevertheless, it is able to capture the main properties of the Peregrine. 
	By} integrating out the transverse degrees of freedom, 
the 3D Gross-Pitaevskii Eqs.~(1) 
reduce~\cite{Pitaevskii2003,Pethick2008,Kevrekidis2015} to 
\begin{align}
	\label{eq:1D_CGPE}
	i\hbar\partial_t\Psi_F(x,t) =
	& \Big[-\frac{\hbar^2}{2m}\partial^2_x + V(x) + V_G(x) \nonumber 
	\\ 
	& + \sum_{F'=1}^2 g_{FF'}^{(1{\rm D})}|\Psi_{F'}(x,t)|^2\Big]\Psi_F(x,t) \,.
\end{align}
Here, $\Psi_F(x,t)$ denote the longitudinal wave functions normalized to the number of particles $N_F = \int |\Psi_F|^2 \dd x$, with $F=1,2$ signifying
the appropriate hyperfine level.
The 1D coupling strength $g_{FF'}^{(1{\rm D})}=2\hbar\omega_\perp a_{FF'}$ refers to the intra- $(F=F')$ and interspecies ($F\neq F'$) interaction, while $a_{FF'}$ is the 3D s-wave scattering length among the corresponding hyperfine states, $F$ and $F'$.  
Motivated by the experimental setup we use herein the transverse trapping frequency $\omega_\perp = 2\pi\times251$~Hz (see also Sec.~\ref{sec:sup1}). 
Moreover, the external trapping potential {is $V(x) = m(\omega/\omega_\perp)^2x^2/2$, with $\omega$ representing  the trap frequency in the $x$-direction,} and the optically induced Gaussian well in 1D reads $V_G(x) = -V_0\exp(-2x^2/s_x^2)$, where $V_0$ ($s_x$) refers to its amplitude (width).

Following the experimental procedure to initialize the dynamics, we first obtain the ground state of the $\Psi_1$ hyperfine state, by means of imaginary time propagation. 
Simultaneously, we ensure that the Thomas-Fermi radius of the condensate is in agreement with that of the experiment, fixing this way 
the total number of particles in this 1D setup.
Next, the following unitary transformation
\begin{align}
	\label{eq:unitary_transformation}
	\mqty(\Psi_1 \\ \Psi_2) = \exp\qty[-i\frac{\pi}{p_0}\mqty(0 & 1 \\ 1 & 0)] \mqty(\Psi_1 \\ \Psi_2) \,,
\end{align}
is applied on the two-component wave function  to mimic the instantaneous transfer atoms from $\Psi_1$ to $\Psi_2$, according to the experimental protocol (see Sec.~II). 
In the above expression the parameter $p_0$ allows to adjust the percentage of transferred atoms. 
For instance, $p_0=4, 8, 16$ corresponds to 50\%, $\sim$15\%, $\sim$4\% population transfer, respectively.  

%--------------------------------------------------------------------------------------------------------
\subsection{Seeding the Peregrine soliton through a Gaussian-well} \label{sec:well}

\begin{figure}[t]
	\centering
	\includegraphics[width=\linewidth]{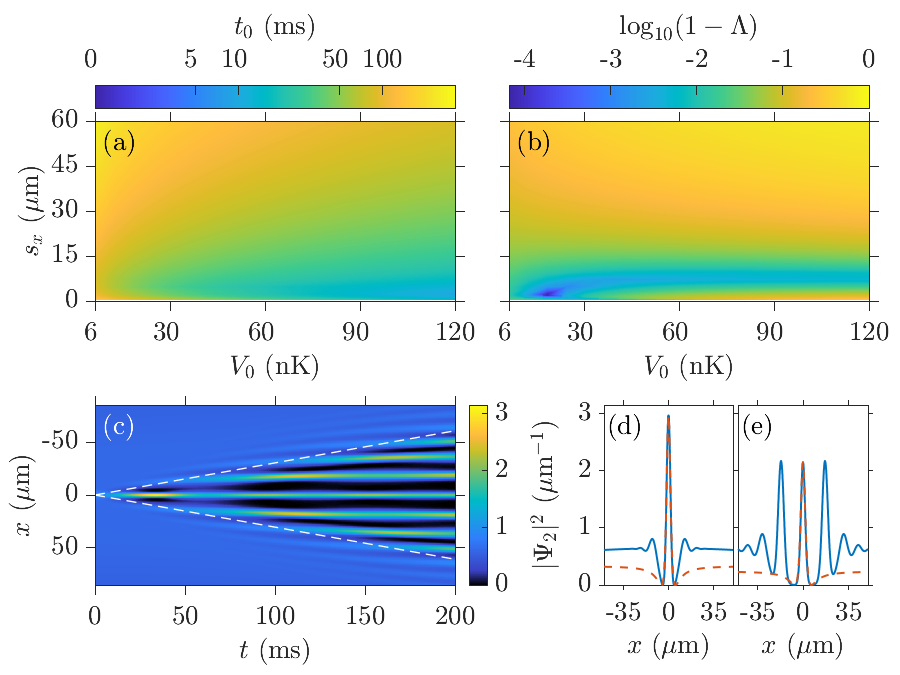}
	\caption{
		(a) Time instant of the emergence of the Peregrine soliton with respect to the width, $s_x$, and depth, $V_0$, of the Gaussian-well. 
		(b) Logarithm of the spatial difference, $\log_{10}(1-\Lambda)$ [see Eq.~\eqref{eq:overlap}], between the numerically obtained structure and the analytic Peregrine solution in the ($s_x$-$V_0$) plane. 
		(c) Evolution of the density of the minority component in the presence of a Gaussian well with $V_0=30$~nK
		and $s_x=4.8~\mu$m.
		White dashed lines mark the universal envelope of the nonlinear stage of the modulational instability Eq.~\eqref{eq:universal_envelope}.
		(d) Numerically identified Peregrine soliton profile (solid blue line) compared to the analytical solution (dashed red line) at $t=33.6$~ms. 
		(e) Formation of three solitonic entities at $t=94.9$~ms. 
		Note the qualitative agreement of the 1D configurations with the ones observed by the experiment and the 3D simulations shown in Fig.~1 of the main text.
	}
	\label{fig:landscape_well}
\end{figure}

While the methodology proposed in Ref.~\cite{Romero-Ros2022a} is, in principle, suitable for producing a Peregrine soliton, the long appearance times of the latter and the limited controllability 
of the scheme prompted us to devise {\it key} structural modifications in the experiment herein including, notably, the inclusion of the Gaussian potential well.

Relying on the 1D mean-field description [Eq.~\eqref{eq:1D_CGPE}] we unravel the emergence of the Peregrine wave as a function of the characteristics of the Gaussian well within the range $V_0\in[6, 120]$~nK
and $s_x\in[2.4, 96.3]~\mu$m.
Specifically, the system is allowed to evolve up to $\sim 320$~ms and the time instant of peak formation, 
$t_0$, 
as well as the spatial overlap, $\Lambda$, between  
the numerically obtained structure and the analytic Peregrine solution (2) are estimated [see Fig.~\ref{fig:landscape_well}].  
The underlying spatial overlap~\cite{Jain2011,Mistakidis2018} is defined as,
\begin{equation}
	\label{eq:overlap}
	\Lambda(t) = \frac{\qty[\int\dd x |\Psi_2|^2|\Psi_P|^2]^2}{\int\dd x|\Psi_2|^4\int\dd x|\Psi_P|^4} \,,
\end{equation}
where $\Psi_2(x,t_0)$ corresponds to the emergent structure at $t_0$ and $\Psi_P(x)$ is the analytical Peregrine solution 
obtained as described in the main text [see the relevant discussion around Fig.~3].
The integration limits are taken at the points where the Peregrine wave function presents a phase jump of $\pi$, i.e., when $|\Psi_P(x)|^2=0$, corresponding to $x=\pm\sqrt{3}L_P/2$.
If $\Lambda=1$, the core of the numerically identified entity is identical to that of the analytical prediction. 

It is found that both $s_x$ and $V_0$ facilitate the controllable Peregrine nucleation. 
Particularly, a close inspection of Fig.~\ref{fig:landscape_well}(a) reveals that by increasing $s_x$ of the Gaussian-well, the 
Peregrine formation is delayed. 
Note here that the impact of the width variation is more pronounced for shallower wells.
{
	On the other hand, in Fig.~\ref{fig:landscape_well}(b) we depict the logarithm of the \textit{spatial difference}, $\log_{10}(1-\Lambda)$.
	Below $s_x=30~\mu$m, the deviation from the analytical Peregrine (2) is less than a 10\%.
	More specifically, we found that the optimal $s_x$ is around the charateristic length scale of the Peregrine soliton, given by $L_p=\sqrt{\hbar^2/m|g_{\rm eff}|P_0}\approx 6.7~\mu$m.
	Moreover, this value is independent of $V_0>30$~nK, as depicted by the blue region in Fig.~\ref{fig:landscape_well}(b).
	As $s_x\to0$ or $s_x\gg L_p$, we find that $\Lambda < 0.9$ and we cannot attribute to the emergent peak structures a definitive Peregrine character.
}

These results are in line with the experimental findings reported in Fig.~1  of the main text. 
The spatiotemporal evolution of the minority component in the presence of a Gaussian well with $V_0  = 30$~nK
and $s_x=4.8~\mu$m is presented in Fig.~\ref{fig:landscape_well}(c). 
Generation of the Peregrine soliton occurs at $t=33.6$~ms [see solid line in Fig.~\ref{fig:landscape_well}(d)]. 
The analytical Peregrine solution~(2) is also provided for comparison  showcasing an excellent agreement.
Notice here, that the triplet structure reported by the experiment and captured by 3D mean-field simulations emerges also in the 1D setting at $t = 94.9$~ms [Fig.~\ref{fig:landscape_well}(e)]. 

Finally, despite being outside of the main scope of the present work, we remark that we came across the nonlinear stage of modulational instability~\cite{El1993,Biondini2016b,Zhao2016}. 
This phenomenon appears in purely focusing media being characterised by a universal envelope with boundaries~\cite{Zakharov2013,Biondini2018a,Kraych2019}
\begin{equation}
	\label{eq:universal_envelope}
	x_\pm = \pm 2\sqrt{-2g_{{\rm eff}}P_0}t \,. 
\end{equation}
These are illustrated in Fig.~\ref{fig:landscape_well}(c) with white dashed lines.  
In this expression, $g_{{\rm eff}}=g^{(1{\rm D})}_{22}-\qty(g^{(1{\rm D})}_{12})^2/g_{11}^{(1{\rm D})}<0$ is the effective interaction strength of the respective single-component reduction~\cite{Dutton2005,Bakkali-Hassani2021,Romero-Ros2022a} and $P_0$ is the amplitude of the background. 
The latter is also provided by the peak of the Peregrine soliton [see e.g. Fig.~\ref{fig:landscape_well}(d) and the discussion in the main text around Fig.~3].

In this regard, a key feature of the present work
(in comparison to the earlier findings
and proposal of Ref.~\cite{Romero-Ros2022a}), is that the usage of a Gaussian well triggers the modulational instability at earlier times of the dynamics.
This, in turn, paves a new way to study in a controlled environment, and more importantly,  in experimentally accessible time scales, the emergence of such phenomena.

\end{document}